\documentclass[twocolumn,twoside,slac_two]{revtex4}
\usepackage{graphicx}
\usepackage{fancyhdr}
\usepackage[colorlinks=true,urlcolor=magenta]{hyperref}
\newcommand{\dif}{\mathrm{d}}
\newcommand{\nvec}{\mathbf{n}}

\begin{document}

\title{Cosmic-Ray Anisotropies: A Review}


\author{O. Deligny}
\affiliation{Institut de Physique Nucl\'eaire, CNRS-IN2P3, Univ. Paris-Sud, Universit\'e Paris-Saclay, 91406 Orsay Cedex, France}

\begin{abstract}
Important observational results have been recently reported on the angular distributions of cosmic rays (CRs) at all energies, calling into question the perception of CRs a decade ago. These results together with their in-progress interpretations are summarized in this short overview paper, following the contribution presented at the    {\MakeUppercase{\romannumeral 25}} European Cosmic Ray Symposium.
\end{abstract}

\maketitle

\thispagestyle{fancy}


\section{Introduction}

The origin of CRs remains an enduring question in astrophysics. The arrival directions of these particles are highly isotropic. This is expected from the propagation of charged particles in the interstellar medium where the directions of the particle momenta are randomized over time by the effective scattering in the encountered magnetic fields. Due to the scrambling action of these fields,  small anisotropy contrats are expected to be imprinted upon the distribution of arrival directions of CRs as observed on Earth. Only for the population of ultra-high energy CRs, magnetic deflections could be small enough to allow for mirroring to some extent the distribution of sources and the observed patterns, but the small flux combined to the potential absence of particles with low electric charge at these energies still prevents such a 'charged-particle astronomy' with current data.  

During the past decade, multiple observatories located in both hemispheres have reported significant  observations of large-scale and small-scale anisotropies in the TeV-PeV energy band. These results have challenged the long-standing description of CR propagation in terms of a typical spatial diffusion process from stationary sources located preferentially in the disk of the Galaxy, leading to a dipole moment only in the direction of the CR gradient and with an amplitude steadily increasing with the energy. The current picture, much more complex and elaborated, is discussed in section~\ref{sec:tev-pev}. 

In the PeV-EeV energy range, the possible increase of anisotropy amplitudes does not allow, yet, for compensating the decrease of the flux and of the collected statistics. There are essentially upper limits on the amplitude and phase measurements for the first harmonic in right ascension, which are presented in section~\ref{sec:pev-eev}. 

At EeV energies and above, the study of ultra-high energy CRs has experienced a jump in statistics as well as improved instrumentation in the past decade. This has allowed a better sensitivity in searching for anisotropies. Currently, the stronger constraints come from measuring the first harmonic modulation in right ascension. At the highest energies, searches for clustering of arrival directions as well as searches for correlations between CR arrival directions and the positions of nearby extragalactic objects are the best suited tools in attempting to decipher the origin of these particles. These searches are summarized in section~\ref{sec:eev-zev}. 

Prior to reviewing these results, and since most of the analysis techniques aim at extracting the moments of the observed angular distributions to characterize the anisotropies, general reminders on the formalism of the moment reconstruction are first provided in section~\ref{sec:analysis}.

\section{Extracting the Moments of the Angular Distributions}\label{sec:analysis}

\subsection{Harmonic Analysis in Right Ascension}

At energies $\gtrsim 10~$TeV, measurements of CRs are indirect due to too low primary fluxes and are usually performed through extensive air shower arrays. These arrays operate almost uniformly with respect to sidereal time thanks to the rotation of the Earth : the zenith-angle-dependent shower detection is then a function of declination but not a function of right ascension. Thus, the most commonly used technique is the analysis in right ascension only, through harmonic analysis of the counting rate within the declination band defined by the detector field of view. Considered as a function of the right ascension only, the flux of CRs can be decomposed in terms of a harmonic expansion:
\begin{equation}
\label{eqn:phi}
\Phi(\alpha)=a_0+\sum_{n>0}~a_n^c~\cos{n\alpha}+\sum_{n>0}~a_n^s~\sin{n\alpha}.
\end{equation}
The customary recipe to extract each harmonic coefficient makes use of the orthogonality of the trigonometric functions~\cite{Linsley1975}. Modelling any observed arrival direction distribution, $\overline{\Phi}(\alpha)$, as a sum of $N$ Dirac functions over the circle, $\overline{\Phi}(\alpha)=\sum_i \delta(\alpha,\alpha_i)$, the coefficients can be estimated from the discrete sums:
\begin{equation}
\label{eqn:an1}
\overline{a}_n^c=\frac{2}{N} \sum_{1\leq i \leq N} \cos{n\alpha_i},\,\,\,
\overline{a}_n^s=\frac{2}{N} \sum_{1\leq i \leq N} \sin{n\alpha_i}.
\end{equation}
Here, the re-calibrated harmonic coefficients $a_n^c\equiv a_n^c/a_0$ and $a_n^s\equiv a_n^s/a_0$ are directly considered, as it is traditionally the case in measuring \textit{relative} anisotropies. In addition, over-lined symbols are used to indicate the \textit{estimator} of any quantity. The statistical properties of the estimators $\{\overline{a}_n^c,\overline{a}_n^s\}$ can be derived from the Poissonian nature of the sampling of $N$ points over the circle distributed according to the underlying angular distribution $\Phi(\alpha)$. In case of small anisotropies (\textit{i.e.} $|a_n^c/a_0|\ll 1$ and $|a_n^s/a_0|\ll 1$), the harmonic coefficients are recovered with an uncertainty such that $\sigma_n^c(\overline{a}_n^c)=\sigma_n^s(\overline{a}_n^s)=\sqrt{2/N}$. For an isotropic realization, $\overline{a}_n^c$ and $\overline{a}_n^s$ are random variables whose joint p.d.f., $p_{A_n^c,A_n^s}$, can be factorized in the limit of large number of events in terms of two Gaussian distributions whose variances are thus $\sigma^2=2/N$. For any $n$, the joint p.d.f. of the estimated amplitude, $\overline{r}_n=(\overline{a}_n^{c2}+\overline{a}_n^{s2})^{1/2}$, and phase, $\overline{\phi}_n=\arctan{(\overline{a}_n^s/\overline{a}_n^c)}$, is then obtained through the Jacobian transformation~:
\begin{eqnarray}
\label{eqn:jointpdf1}
p_{R_n,\Phi_n}(\overline{r}_n,\overline{\phi}_n)=\frac{\overline{r}_n}{2\pi\sigma^2}~\exp{(-\overline{r}_n^2/2\sigma^2)}.
\end{eqnarray}
From this expression, it is straightforward to recover the Rayleigh distribution for the p.d.f. of the amplitude, $p_{R_n}$, and the uniform distribution between 0 and $2\pi$ for the p.d.f. of the phase, $p_{\Phi_n}$. Overall, this formalism provides the amplitude of the different harmonics, the corresponding phase (right ascension of the maximum intensity), and the probability of detecting a signal due to fluctuations of an isotropic distribution with an amplitude equal or larger than the observed one as $P(>\overline{r}_n)=\exp{(-N\overline{r}_n^2/4)}$. 

Note that the formalism aforementioned can be applied off the shelf only in the case of a purely uniform directional exposure, condition which is generally not fulfilled. At the \textit{sidereal} time scale, the directional exposure of most observatories operating with high duty cycle (\textit{e.g.} surface detector arrays) is however only moderately non-uniform. Different approaches are then adopted in the literature, and the non-uniform directional exposures presented in next sections are carefully accounted for.

\subsection{Multipole Expansion in Right Ascension and Declination}

In general, and in contrast to the simplified approach presented in the last subsection, the flux of CRs $\Phi(\nvec)$ can depend on both the right ascension and the declination and thus be decomposed in terms of a multipolar expansion onto the spherical harmonics $Y_{\ell m}(\nvec)$:
\begin{equation}
\label{eqn:almexpansion}
\Phi(\nvec)=\sum_{\ell\geq0}\sum_{m=-\ell}^\ell a_{\ell m}Y_{\ell m}(\nvec).
\end{equation}
Non-zero amplitudes in the $\ell$ modes arise from variations of the flux on an angular scale ${\simeq}1/\ell$ radians. The partial-sky coverage of ground-based observatories prevents the multipolar moments $a_{\ell m}$ to be recovered in a direct way through the customary recipe making use of the orthogonality of the spherical harmonics basis~\cite{Sommers2001}. Indirect procedures have to be used, one of them consisting in considering first the 'pseudo'-multipolar moments
\begin{equation}
\tilde{a}_{\ell m} = \int \dif\mathbf{n}~\omega(\mathbf{n})\Phi(\mathbf{n})Y_{\ell m}(\mathbf{n}),
\end{equation}
and then the system of linear relationships relating these pseudo moments to the real ones:
\begin{equation}
\tilde{a}_{\ell m} = \sum_{\ell'\geq0}\sum_{m'=-\ell}^\ell a_{\ell' m'}\int \dif\mathbf{n}~\omega(\mathbf{n})Y_{\ell m}(\mathbf{n})Y_{\ell' m'}(\mathbf{n}).
\end{equation}
Assuming a bound $\ell_{\mathrm{max}}$ beyond which $a_{\ell m}=0$, these relations can be inverted allowing the recovering of the moments $a_{\ell m}$. However, the obtained resolution on each moment does not behave as $\sqrt{A/N}$ (with $A$ a constant factor depending on $\omega$) as expected from naive statistical arguments, but increases exponentially with $\ell_{\mathrm{max}}$~\cite{Billoir2008}.

It is to be noted that, in many practical cases where the declination dependence of the directional exposure $\omega(\alpha,\delta)$ is too difficult to estimate down to the required accuracy, analyses are performed by reproducing the observed declination dependence of the event counting rate. In such cases, anisotropies are captured in right ascension only, be it in several declination bands.

\subsection{Angular Power Spectrum}

Hence, in general, the small values of the energy-dependent $a_{\ell m}$ coefficients combined to the available statistics in the different energy ranges does not allow for an estimation of the individual coefficients with a relevant resolution as soon as $\ell_{\mathrm{max}}>2$. However, based on analysis techniques previously developed in the CMB community, it is possible, under some restrictions detailed and discussed in~\cite{Auger2016a}, to reconstruct the angular power spectrum coefficients ${C}_{\ell}=\sum^{\ell}_{m=-\ell}|a_{\ell m}|^{2}/(2\ell +1)$ within a statistical resolution independent of the bound $\ell_{\mathrm{max}}$. The starting point is to consider any observed distribution of arrival directions as a particular realization of an underlying Gaussian process. The simplest non-trivial situation to describe the underlying process is then to consider that the anisotropies cancel in ensemble average and produce a second order moment that does not depend on the position on the sphere but only on the angular separation between $\mathbf{n}$ and $\mathbf{n'}$. In this case, the underlying $a_{\ell m}$ coefficients vanish in average and are not correlated to each other (\textit{i.e.} diagonal covariance: $\langle a_{\ell m}a_{\ell' m'}\rangle=C_\ell\delta_{\ell\ell'}\delta_{mm'}$) so that the ${C}_{\ell}$ coefficients can be viewed as a measure of the variance of the $a_{\ell m}$ coefficients. In this situation, it can then be shown that the pseudo-power spectrum $\tilde{C}_{\ell}=\sum^{\ell}_{m=-\ell}|\tilde{a}_{\ell m}|^{2}/(2\ell +1)$ (which is directly measurable) is related to the real power spectrum through
\begin{eqnarray}
\label{eqn:Cl}
\tilde{C}_{\ell} = \sum_{\ell'}M_{\ell\ell'}C_{\ell'},
\end{eqnarray}
where the operator $M$ describing the cross-talk induced by the non-uniform exposure between genuine modes is entirely determined by the knowledge of the exposure function. Various techniques are then available to estimate the ${C}_{\ell}$ coefficients from the measured $\tilde{C}_{\ell}$ ones. 

Caution should be kept in mind when interpreting angular power spectra measured with partial-sky coverages: invoking a random nature for the arrival direction distributions of CRs allows for a stochastic modelling of non-well known or unknown source positions and propagation mediums; in opposition to CMB studies where the power spectrum results from primordial and fundamental fluctuations. As a result, the angular distributions in the uncovered regions of the sky could show patterns that would lead to different power spectra.

\section{Large-Scale and Small-Scale Anisotropies in the TeV$-$PeV Energy Range}\label{sec:tev-pev}

\subsection{First Harmonic Measurements and Implications}

\begin{figure}[!ht]
  \centering
  \includegraphics[width=7.5cm]{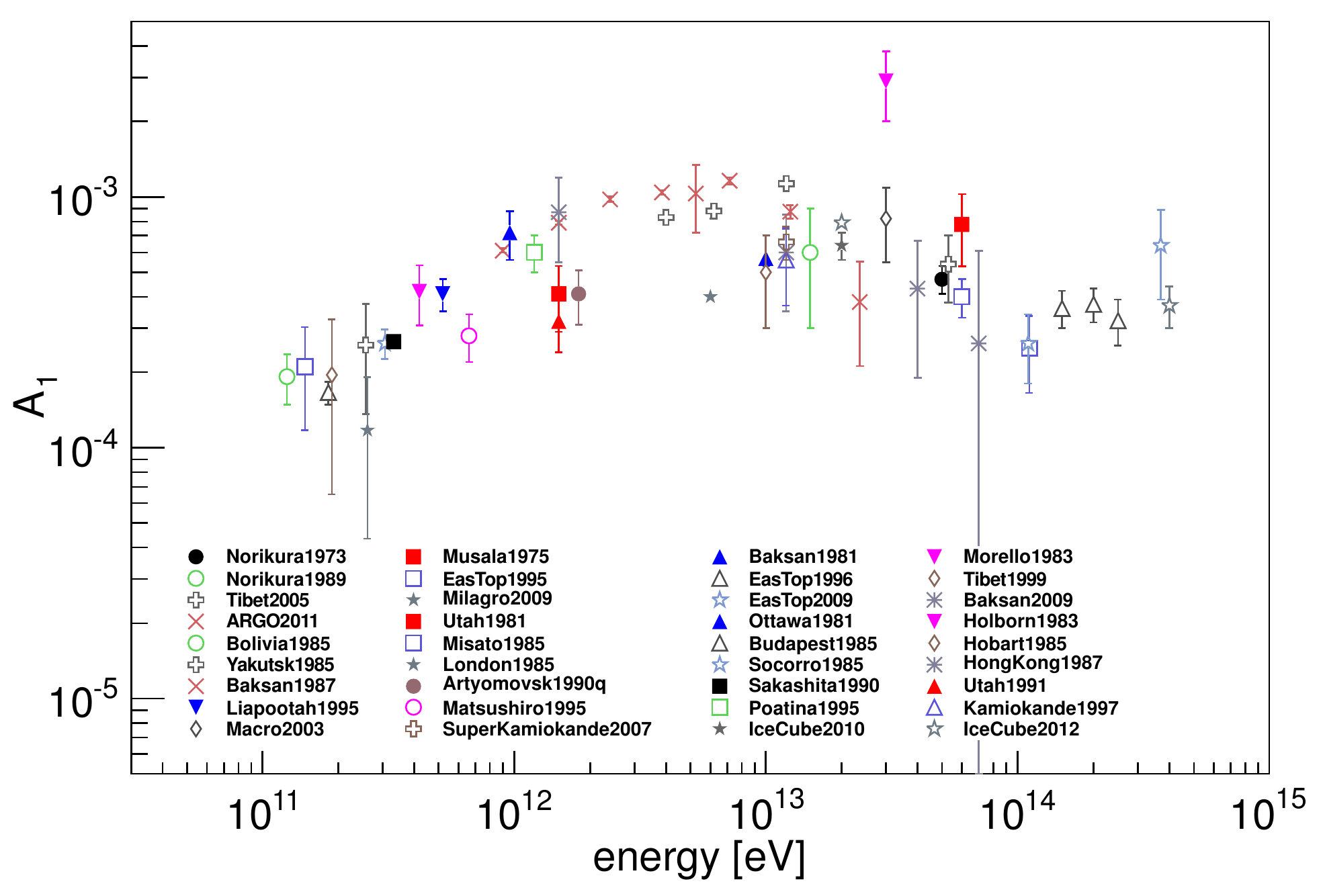}
  \includegraphics[width=7.5cm]{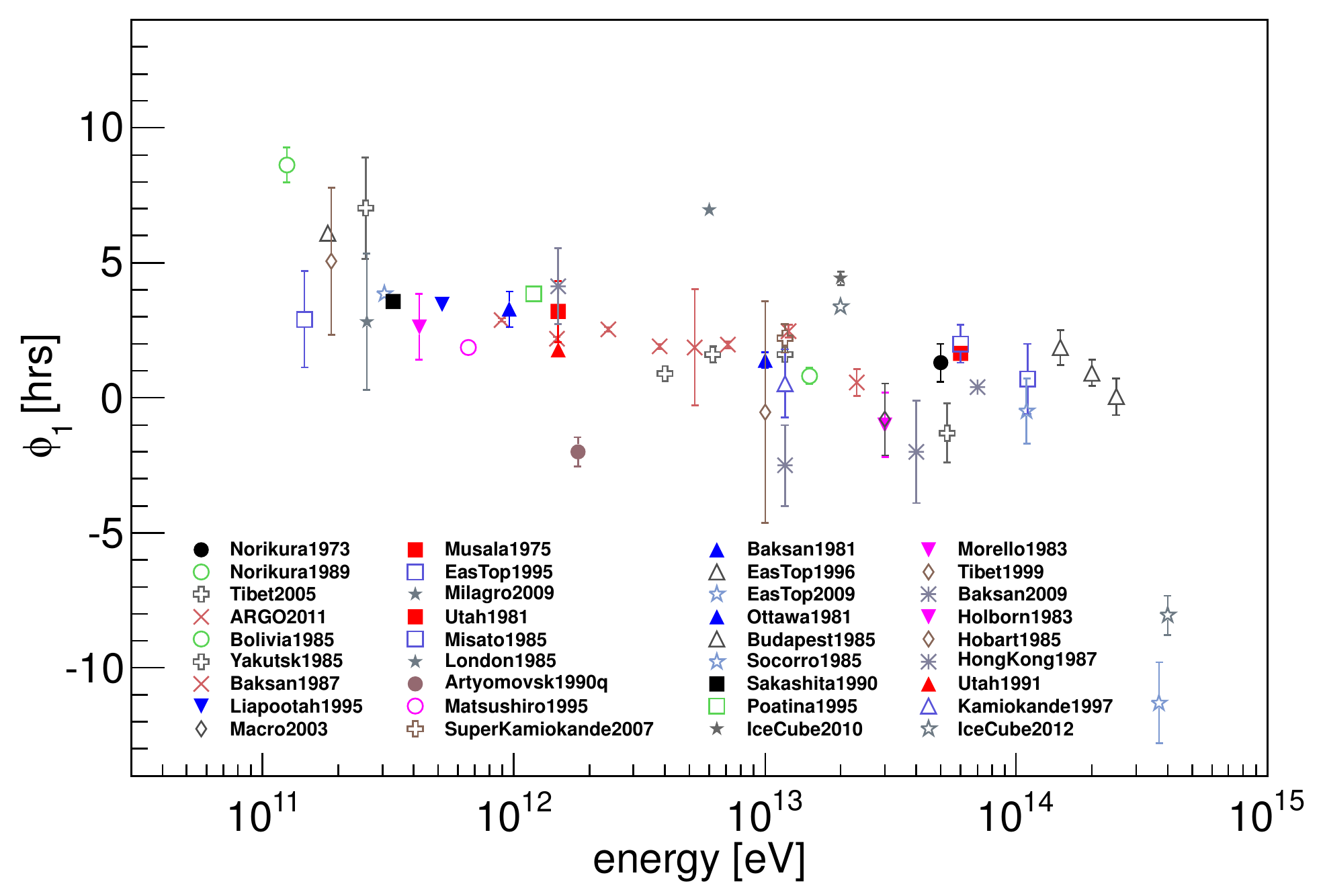}
  \caption{First harmonic amplitude (top) and phase (bottom) of the sidereal daily variations as a function of the energy measured by underground muon detectors and extensive air shower arrays, as collected in~\cite{DiSciascio2013}.}
  \label{fig:ampphase-tev}
\end{figure}

In the TeV-PeV energy range, complex patterns have been revealed in the arrival directions of CRs thanks to the large statistics collected in the last decade by several experiments, and anisotropy contrats at the $10^{-4}-10^{-3}$ level are now established at large scales. Consistent measurements from experiments located in both hemispheres were reported: Tibet AS$\gamma$~\cite{Tibet2005},  Super-Kamiokande~\cite{SK2007},  Milagro~\cite{Milagro2008}, EAS-TOP~\cite{EASTOP2009}, MINOS~\cite{MINOS2011}, ARGO-YBJ~\cite{ARGO2013},  and HAWC~\cite{HAWC2013} in the Northern hemisphere, and IceCube~\cite{IceCube2010} and IceTop~\cite{IceTop2012} in the Southern hemisphere. A collection of amplitude and phase measurements from~\cite{DiSciascio2013} is shown in figure~\ref{fig:ampphase-tev} including a series of former experiment results: the amplitude is observed to increase with energy up to $\simeq 10~$TeV before flattening, and the phase is observed to be smoothly evolving before undergoing a sudden flip at $\simeq 0.3~$PeV. Note that small corrections to the amplitudes related to the different latitudes and zenithal-dependent efficiencies of the different experiments are not applied here. 

The first harmonic parameters as derived in figure~\ref{fig:ampphase-tev} are generally considered with special interest due to their relationships with the dipole moment. In the context of spatial diffusion, the dipole moment is naively expected to provide a way to probe the particle density gradient shaped by the diffusion in interstellar magnetic fields on scales of the scattering diffusion length. In this picture, for stationary sources smoothly distributed in the Galaxy, the dipole vector should align roughly with the direction of the Galactic center with an amplitude increasing with energy in the same way as the diffusion coefficient, typically $E^{0.3-0.6}$. 

However, this simple picture is not confirmed by the measurements, showing that the dipole amplitude is not described by a single power law and that the dipole phase does not align with the Galactic center and undergoes a rapid flip at an energy of 0.1-0.3 PeV. Recent studies have put this picture into question by considering spatial and temporal stochastic distributions of sources: even for isotropic diffusion governed by a single diffusion coefficient growing as $E^{0.3-0.6}$ with energy, fluctuations of the dipole parameters are then naturally induced by local and young sources even if their contribution to the total flux is only subdominant~\cite{Erlykin2006,Blasi2012,Sveshnikova2013}. On top of the background diffuse emission, the contribution of such sources within some energy intervals can lead, depending on the source population and the diffusion region, to phase rotations and amplitude modulations as a result of the summation of competing dipole vectors associated to each source. 

\begin{figure}[!ht]
  \centering
  \includegraphics[width=5.5cm]{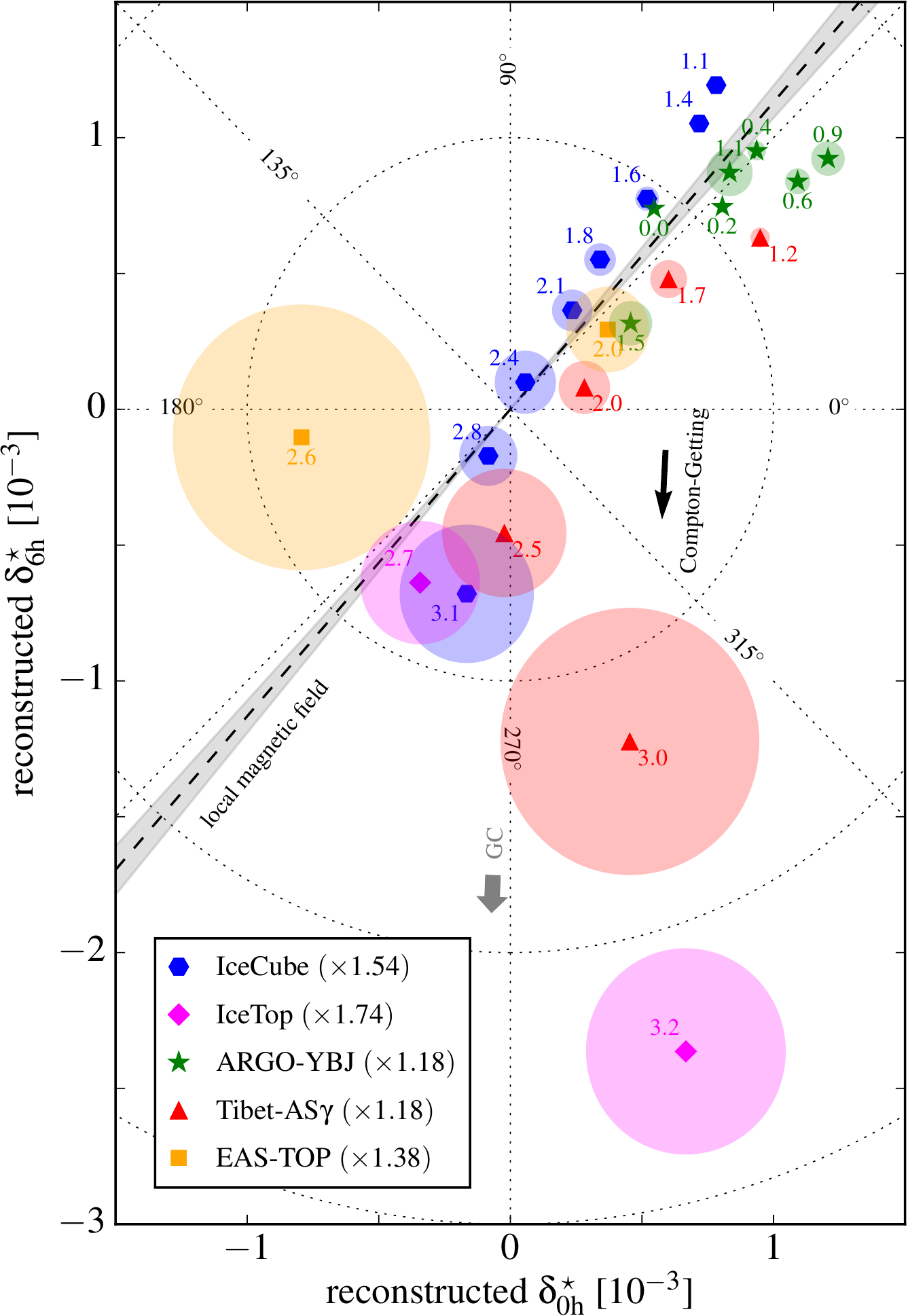}
  \caption{Reconstructed equatorial components of the dipole as a function of energy (median energy indicated next to each data point), after subtraction of the dipole induced by the Compton-Getting effect~\cite{Ahlers2016}.}
  \label{fig:ahlers}
\end{figure}

Within the framework of isotropic diffusion, a phase flip with vanishing amplitude requires, nevertheless, fine tuning of the contribution of local sources on top of the diffuse emission. Additional effects may enter into play to shape the dipole parameters actually observed, such as the anisotropic diffusion induced by the local ordered magnetic field whose strength is such that circular motions predominate over random scattering. In this case, the observed dipole could result from the projection of the density gradient of CRs onto the direction of the ordered magnetic field~\cite{Mertsch2015,Ahlers2016}. Assuming that the measured first harmonic amplitude $r$ and phase in right ascension $\varphi$ are predominantly determined by the dipole vector $\mathbf{d}$, the equatorial dipole components can be derived as $(d_{0\mathrm{h}},d_{6\mathrm{h}})=K(r\cos{\varphi},r\sin{\varphi})$, with $K$ a parameter depending on each experiment and $d_{0\mathrm{h}}$ and $d_{6\mathrm{h}}$ the vector components in the equatorial plane associated to the unit vectors pointing towards local sidereal times 0~h and 6~h. Assuming on the other hand that the local standard of rest (LSR) corresponds to the local plasma frame in which CRs are isotropic, the contribution to the observed dipole components $(d_{0\mathrm{h}},d_{6\mathrm{h}})$ arising from the relative motion of the solar system through the LSR, known as the Compton-Getting effect~\cite{Compton1935}, can be subtracted. Doing so, the summary plot of the reconstructed TeV-PeV dipole components in the equatorial plane obtained recently in~\cite{Ahlers2016} is shown in figure~\ref{fig:ahlers}. The numbers attached to the data
indicate the median energy of the bins as $\log_{10}(E_{\mathrm{med}}/\mathrm{TeV})$. The dashed line and gray-shaded area indicate the magnetic field direction and its uncertainty (projected onto the equatorial plane) inferred from IBEX observations~\cite{IBEX}. A close alignment of the inferred dipole components with the local magnetic field direction is observed. Overall, within this framework of anisotropic diffusion, although the projection of the dipole onto the magnetic field axis does not allow one to reconstruct the CR gradient, the magnetic hemisphere of the CR gradient can be determined by the dipole phase and aligns with Galactic longitudes between 120$^\circ$ and 300$^\circ$ below 0.1-0.3  PeV. In this regard, and among other requirements, the Vela SNR is shown to be a plausible source candidate in~\cite{Ahlers2016}. 

\subsection{Angular Power Spectrum}

\begin{figure}[!ht]
  \centering
  \includegraphics[width=7.5cm]{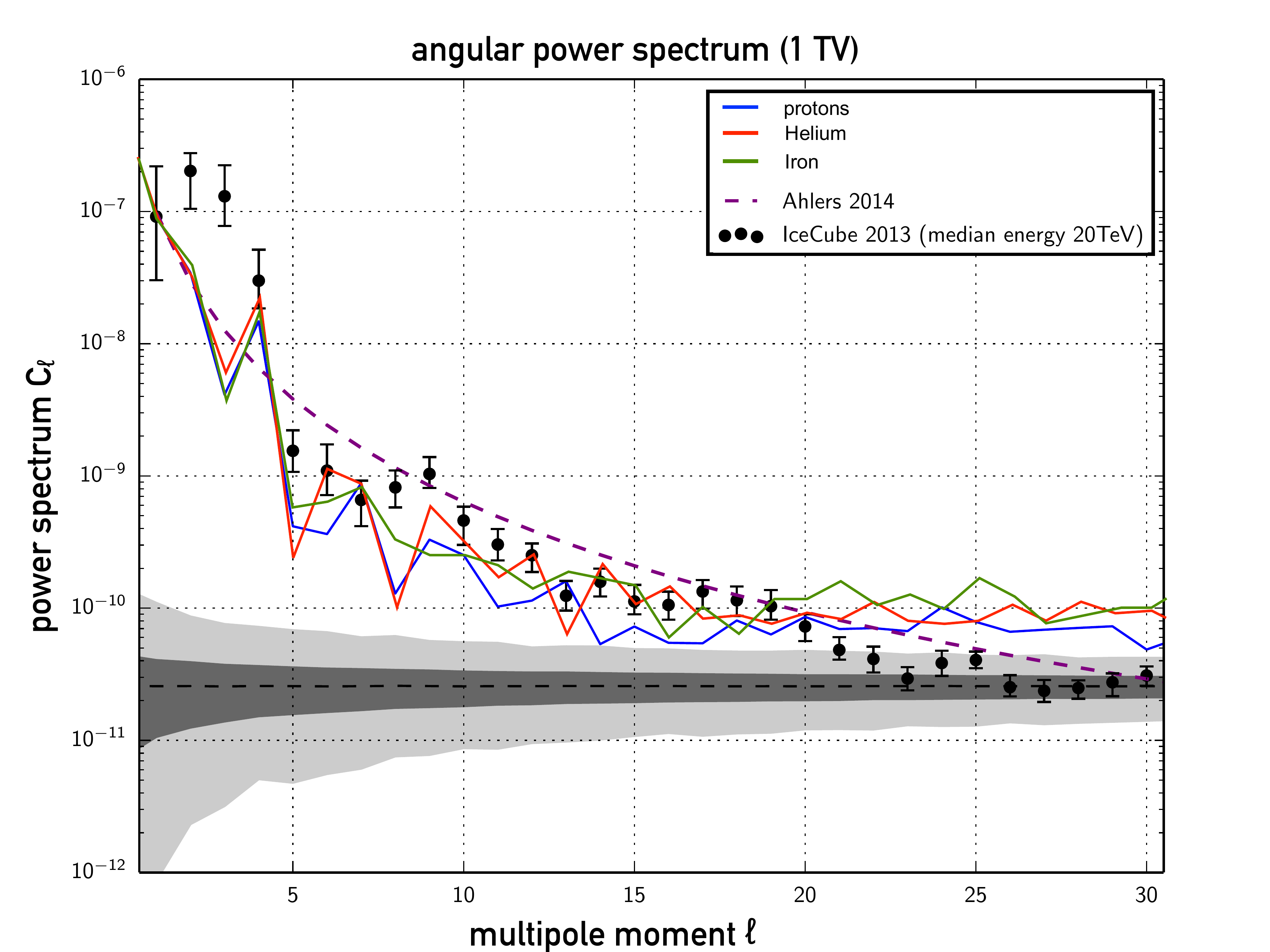}
  \caption{Angular power spectrum as measured by the IceCube collaboration for a median energy of 20~TeV, and model interpretations~\cite{LopezBarquero2016}.}
  \label{fig:powerspectrum}
\end{figure}

Beyond the large-scale anisotropies captured by the first harmonic in right ascension, smaller but significant anisotropy contrasts at the $10^{-5}-10^{-4}$ level at intermediate and smaller scales
have also been reported in the last decade~\cite{Milagro2008,IceCube2011,ARGO2013,HAWC2014,IceCube2016}. The angular power spectrum as derived from IceCube data with a median energy of 20~TeV, shown as the black points in figure~\ref{fig:powerspectrum}, provides a summary plot to visualize the relatively large amount of power at low multipole moments $\ell\leq4$ as well as the small but still significant power at moments up to $\ell=20$. The expected power spectrum fluctuations in 90\% of cases of an isotropic angular distribution is indicated by the light gray band. 

These results have challenged the long-standing picture of CR propagation in terms of a spatial diffusion that would lead predominantly to a dipole moment only. Several interpretations have been put forward. The most conservative one is based on the fact that the observed complex patterns, whose positions appear randomly distributed, are mainly the consequence of the particular structure of the turbulent magnetic field within the last sphere of diffusion encountered by CRs~\cite{Giacinti2012}. With a density gradient of CRs at the entrance of this last sphere of diffusion, the turbulent magnetic field is frozen on the time scale of CR propagation and thus plays the same role as a structured one. The field is then expected to connect regions of higher density outside from the last sphere of diffusion to regions of lower density as seen from Earth, and \textit{vice versa}. This mechanism was shown to produce intermediate- and small-scale anisotropies through Monte-Carlo studies~\cite{Giacinti2012,LopezBarquero2016}, and more recently through an analytical approach~\cite{Ahlers2014,Ahlers2015}. In the former approach, numerical integration of test particles in low $\beta$-compressible magnetohydrodynamic turbulence with the external mean magnetic field as the only free parameter is shown to reproduce reasonably well the power spectrum, as seen in figure~\ref{fig:powerspectrum} through the different solid color lines standing for different primaries~\cite{LopezBarquero2016}. In the latter approach, an initially dipolar distribution is shown to be distorded into a much complex pattern described by higher-order multipoles when introducing second order corrections to the diffusion approximation. These corrections aim at describing the propagation regime between the diffusive particle transport and the deterministic flow of particles. The asymptotic strength of the high-order multipoles at times longer than the diffusion relaxation time is observed to reproduce the data in a reasonable way, as shown by the dotted line in figure~\ref{fig:powerspectrum}~\cite{Ahlers2014}. Overall, thanks to the contemporary measurements of CR arrival directions, the information encompassed in the angular distributions appear today as a tool allowing a possible probe of the local magnetic field environment. 

Besides, it is also to be noted that the power spectrum at small scales measured at TeV energies may also provide relevant signatures to probe and study the electric field induced by the motion of the heliosphere relative to the plasma rest frame where the electromagnetic field can be considered as purely magnetic due to the high conductivity of the medium~\cite{Drury2013}. The anisotropy would then be the consequence of the changes of energy associated to the acceleration/deceleration of the particles, which, given the size and speed of the heliosphere, are expected to produce changes of intensity within small angular scales at the required level for TeV CRs.

\section{Searches for Large-Scale Anisotropies in the PeV$-$EeV Energy Range}\label{sec:pev-eev}

\begin{figure}[!ht]
  \centering
  \includegraphics[width=7.5cm]{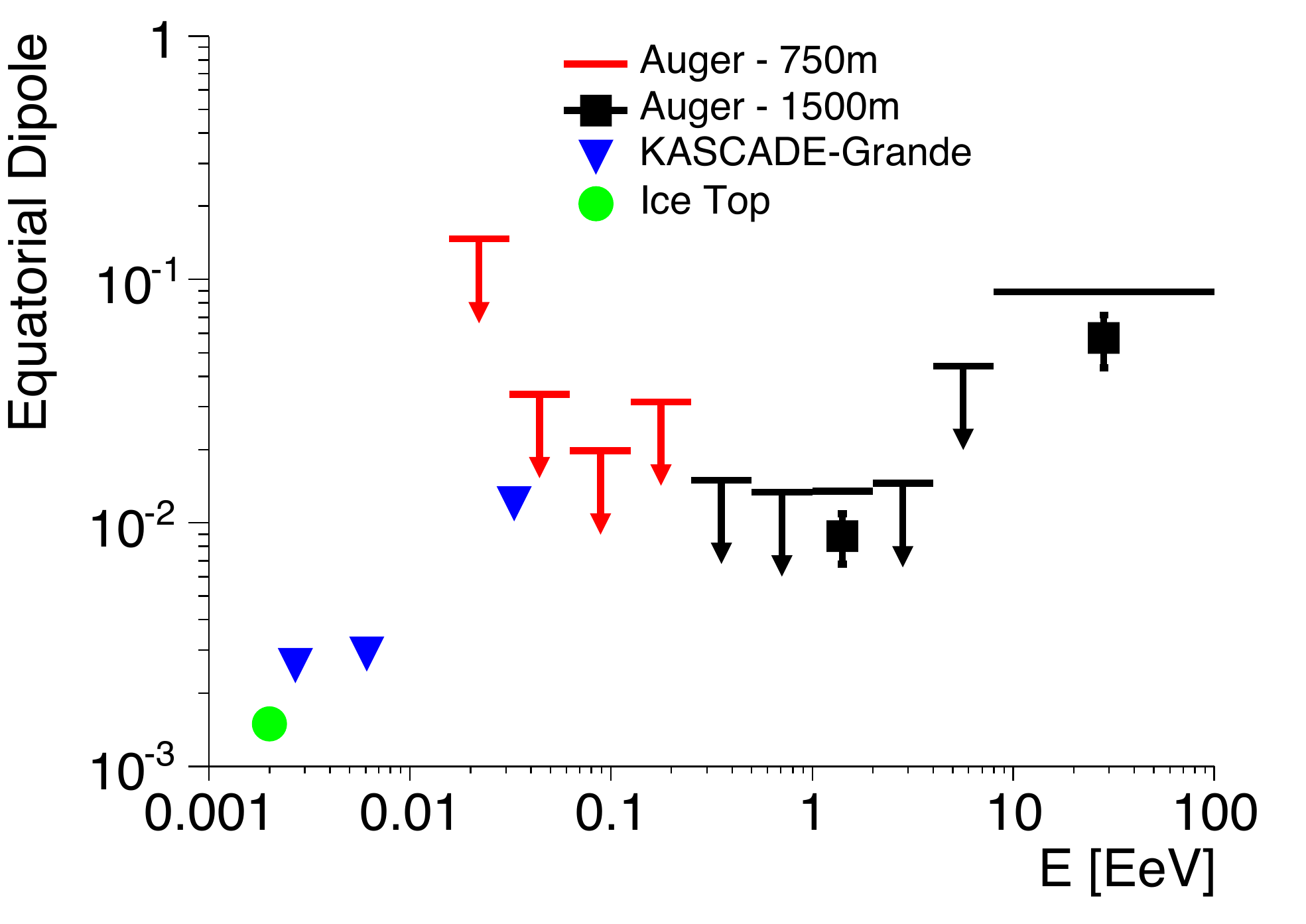}
  \includegraphics[width=7.5cm]{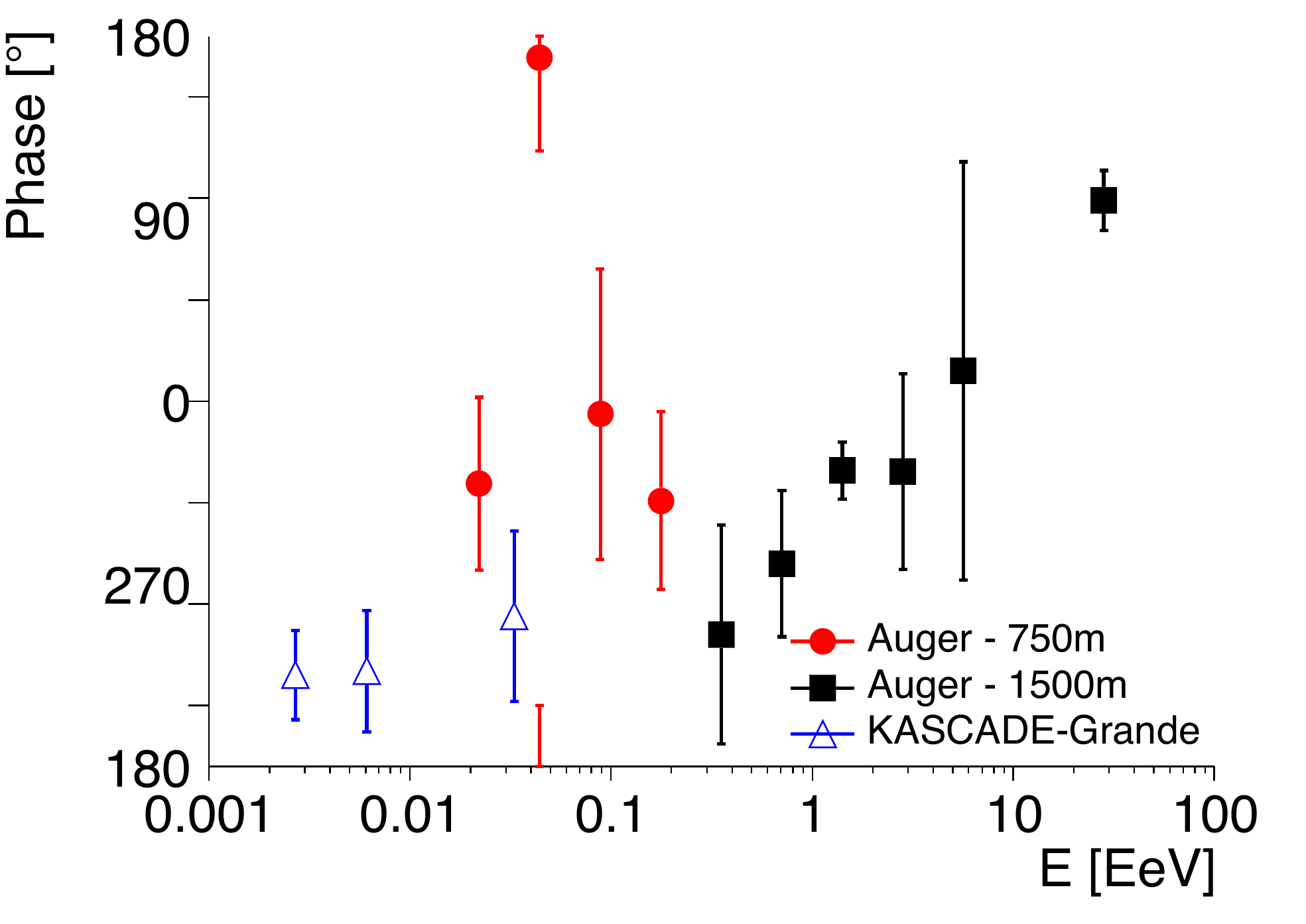}
  \caption{First harmonic amplitude (top) and phase (bottom) as a function of energy above 1~PeV, as collected in~\cite{AlSamarai2015}.}
  \label{fig:peveev}
\end{figure}

At higher energy, in the PeV-EeV energy range, the expected increase of anisotropy contrasts does not compensate, yet, the decrease in the collected statistics with increasing energy. 	The most constraining data, collected in the top panel of figure~\ref{fig:peveev} in terms of the equatorial component of the dipole, are provided by the IceTop~\cite{IceTop2012}, KASCADE-Grande~\cite{KG2015} and Auger~\cite{AlSamarai2015} experiments. Except for the IceTop amplitude, only upper limits are currently provided in this energy range. 

Meanwhile, it is interesting to note that an apparent constancy of phase, even though the significances of the amplitudes are relatively small, has been pointed out previously in surveys of measurements with ion chambers and counter telescope made in the range $\simeq 0.1<E/\mathrm{PeV}<100$~\cite{Greisen1962}. A clear tendency for maxima to occur around 20 hours l.s.t. was observed. As already pointed out by Linsley long time ago, this is potentially indicative of a real underlying anisotropy, because a consistency of the phase measurements in ordered energy intervals is indeed expected to be revealed with a smaller number of events than needed to detect the amplitude with high statistical significance~\cite{Edge1978}. Interestingly, the phases reported by contemporary experiments collected in the bottom panel of figure~\ref{fig:peveev} show a consistent tendency to align in the general right ascension of the Galactic center. 

Above $1~$EeV, a change of phase is observed towards, roughly, the opposite of the one at energies below $1~$EeV. The percent limits to the amplitude of the anisotropy exclude the presence of a large fraction of Galactic protons at EeV energies~\cite{AugerApJS2012,TA2016}. Accounting for the inference from $X_{\mathrm{max}}$ data from both the Pierre Auger Observatory and the Telescope Array that protons are in fact abundant at those energies, this might indicate that this component is extragalactic, gradually taking over a Galactic one. The low level of anisotropy would then be the sum of two vectors with opposite directions, naturally reducing the amplitudes. This scenario is to be explored with additional data.

Increased statistics is thus necessary to probe the anisotropy contrast levels that may exist in this energy range and contain valuable information about the old-age question on the way the transition between Galactic and extragalactic CRs occurs. Also, a current limitation of the measurements is that neither spectra nor anisotropies can yet be studied as a function of the mass of the particles with adequate statistical precision, measurements that would allow a distinction between Galactic and extragalactic angular distributions.

\section{Searches for Anisotropies at the Highest Energies}\label{sec:eev-zev}

\subsection{Large-Scale Anisotropies}

Above $\simeq 10~$EeV, the flux of ultra-high energy CRs (UHECRs) is expected to be of extragalactic origin. Although the actual sources of UHECRs are still to be identified, their distribution in the sky is expected to follow, to some extent, the large-scale structure of the matter in the Universe. It is thus interesting to highlight that in addition to the upper limit on the equatorial dipole component above 8~EeV, the measured amplitude is also shown in figure~\ref{fig:peveev} as the corresponding $p-$value is as low as  $6.4\times10^{-5}$~\cite{Auger2015a}. Assuming that the only significant contribution to the anisotropy is from a dipolar pattern, the amplitude of this signal converts into a $(7.3\pm1.5)\%$ dipole amplitude~\cite{Auger2015a}. This hint may constitute in the near future the first detectable signature of extragalactic CRs observed on Earth.  

To characterize further the angular distribution above 10~EeV, the dipole moment on the sphere is of special interest. An unambiguous measurement of this moment as well as of the full set of spherical harmonic coefficients requires full-sky coverage. Currently, this can be achieved by combining data from observatories located in both the northern and southern hemispheres. To this end, a joint analysis using data recorded at the Pierre Auger Observatory and the Telescope Array above $10$~EeV has been performed in~\cite{AugerTA2014,Deligny2015}. Thanks to the full-sky coverage, the measurement of the dipole moment reported in these studies does not rely on any assumption on the underlying flux of CRs. Note that, in contrast to the results discussed in section~\ref{sec:tev-pev}, the directional exposure derived in these studies is a truly function of both right ascension and declination, so that the reconstructed anisotropy patterns are not absorbed along the declination. 

\begin{figure}[!ht]
  \centering
  \includegraphics[width=7.5cm]{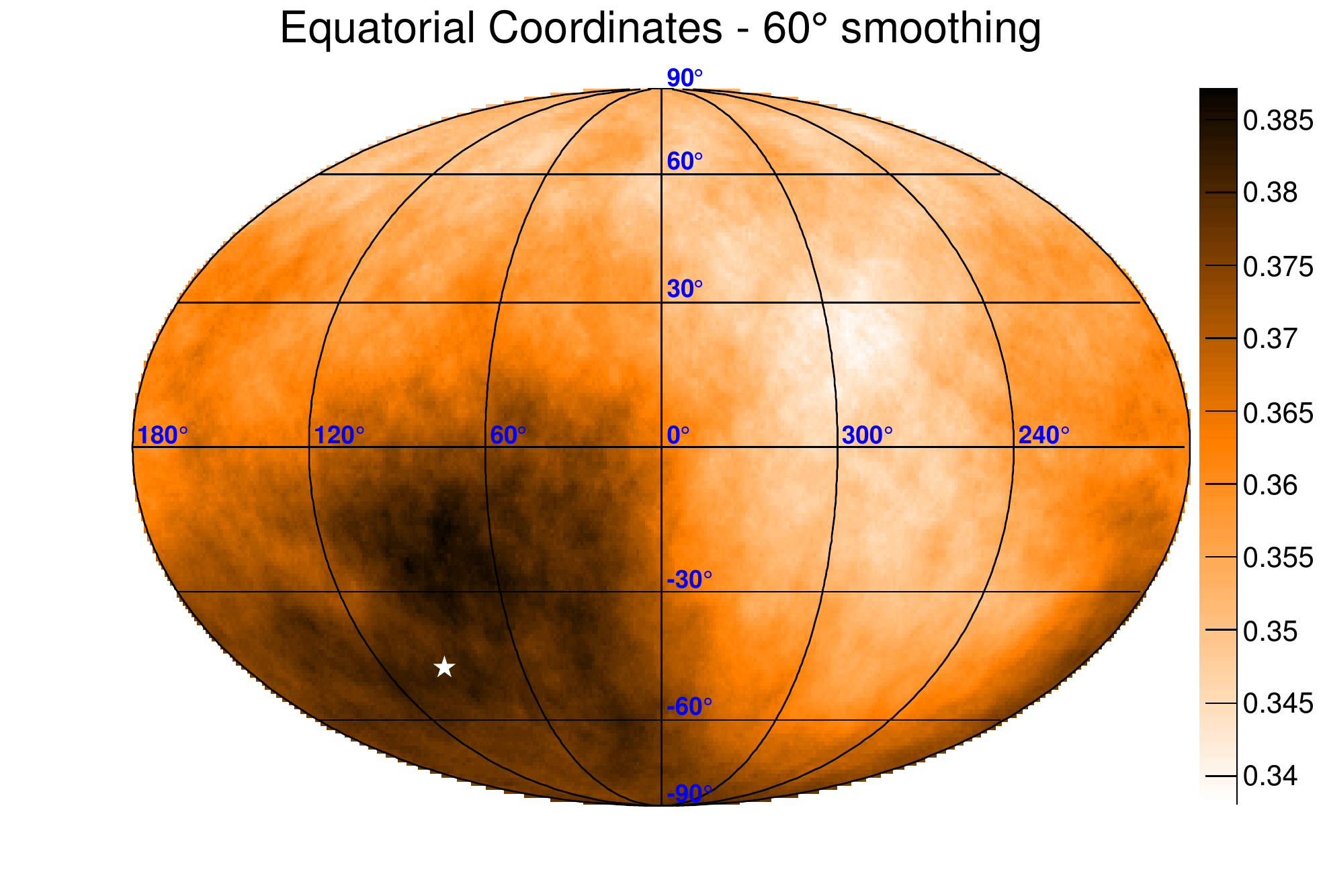}
  \caption{Sky map in equatorial coordinates of the average flux reconstructed from data recorded at the Pierre Auger Observatory and the Telescope Array above $10~$EeV smoothed out at a 60$^\circ$ angular scale, in km$^{-2}$~yr$^{-1}$~sr$^{-1}$ units~\cite{Deligny2015}.}
  \label{fig:augerta}
\end{figure}

The resulting entire mapping of the celestial sphere has revealed a dipole moment with an amplitude $r=(6.5\pm 1.9)\%$, captured with a chance probability of $5\times10^{-3}$. No other deviation from isotropy can be observed at smaller angular scales. The recovered moment can be visualized in figure~\ref{fig:augerta}, where the average flux smoothed out at an angular scale $60^\circ$ per solid angle unit is displayed using the Mollweide projection, in km$^{-2}$yr$^{-1}$sr$^{-1}$
units. This map is drawn in equatorial coordinates. The direction of the reconstructed dipole is shown as the white star.

Large-scale anisotropies of CRs with energies in excess of $10~$EeV are closely connected to the sources and the propagation mode of extragalactic UHECRs, see e.g.~\cite{Harari2014,Tinyakov2015}. Due to scattering in the extragalactic magnetic fields, large deflections are expected even at such high energies for field amplitudes ranging in few nanogauss and extended over coherence lengths of the order of one megaparsec, or even for lower amplitudes if the electric charge of UHECRs is large. For sources distributed in a similar way to the matter in the Universe, the angular distribution of UHECRs is then expected to be influenced by the contribution of nearby sources, so that the Milky Way should be embedded into a density gradient of CRs that should lead to at least a dipole moment. The contribution of nearby sources is even expected to become dominant as the energy of CRs increases due to the reduction of the horizon of UHECRs induced by energy losses more important at higher energies. Once folded through the Galactic magnetic field, the dipole pattern expected from this mechanism is transformed into a more complex structure presumably described by a lower dipole amplitude and higher-order multipoles. However, in these scenarios, the dipole moment could remain the only one at reach within the sensitivity of the current generation of experiments. On the other hand, the detection of significant multipole moments beyond the dipole one could be suggestive of non-diffusive propagation of UHECRs from sources distributed in a non-isotropic way. 

 while it has a small influence

\subsection{Small-Scale Anisotropies}

\begin{figure}[!ht]
  \centering
  \includegraphics[width=7.5cm]{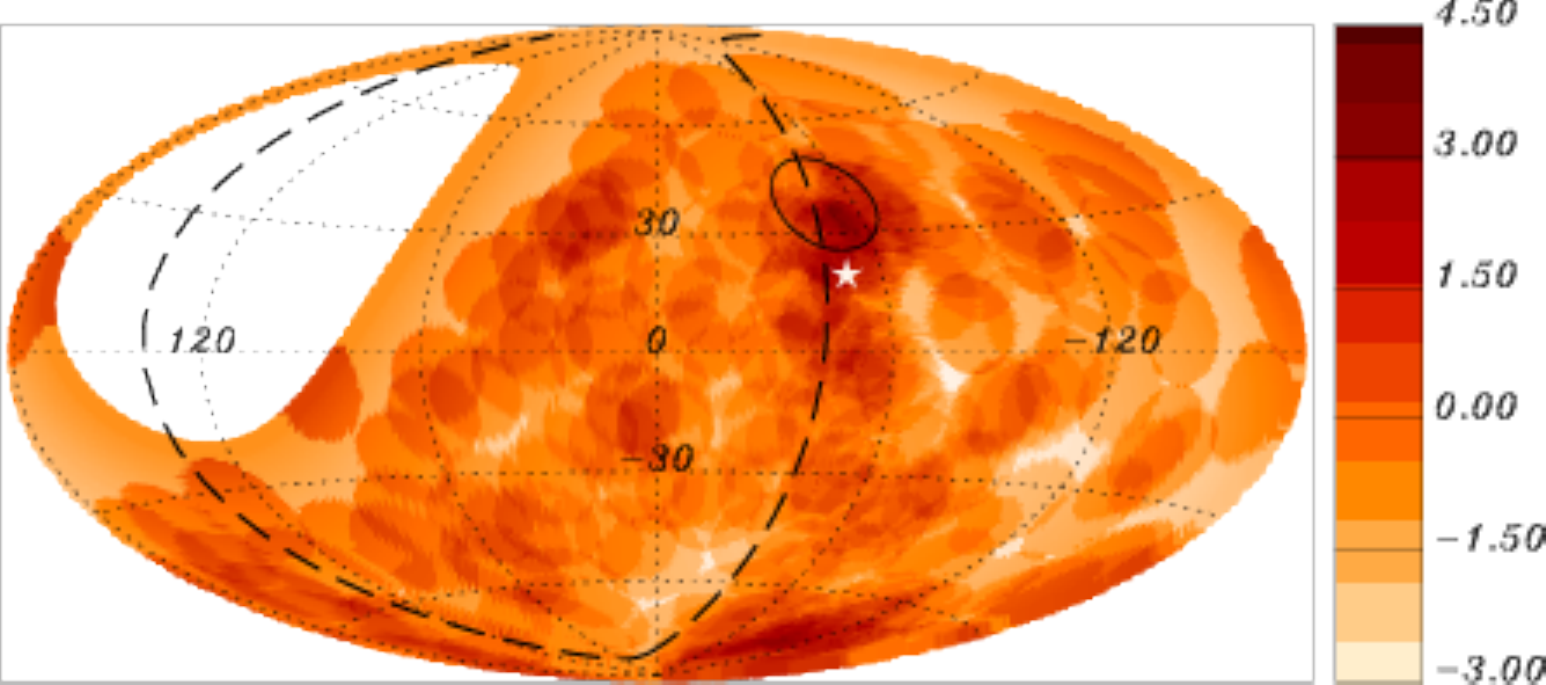}
  \caption{Sky map in Galactic coordinates of the Li-Ma significances of overdensities in 12$^\circ$-radius windows for the Auger events with $E\geq 54$~EeV. Also indicated are the Super-Galactic Plane (dashed line) and Centaurus~A (white star)~\cite{Auger2015}.}
  \label{fig:cenA}
\end{figure}

The energy losses of UHECRs limit the horizon of the highest-energy particles. For small-enough magnetic deflections, the distribution of the arrival directions of UHECRs above $\simeq 40~$EeV could mirror the inhomogeneous distribution of the nearby extragalactic matter. The search for anisotropy at small and intermediate angular scale at the highest energies is thus potentially the most powerful way to infer the sources of UHECRs. 

At the Pierre Auger Observatory, data have been subjected to comprehensive anisotropy searches for different energy thresholds between $40~$ and $80~$EeV, and within different angular windows, between $1^\circ$ and $30^\circ$~\cite{Auger2015}. Searches for significant excesses anywhere in the sky have been performed, as well as searches for correlations with known astrophysical structures and with objects that are considered plausible candidates for UHECR sources. Out of all the searches performed, none of the analyses provides any statistically significant evidence of anisotropy. The two largest departures from isotropy, both with post-trial probability $\simeq1.4\%$, are found for energies in excess of $58~$EeV when looking, on the one hand, within $15^\circ$ of the direction of Centaurus A - the closest radio-loud active galactic nucleus (AGN), and when looking, on the other hand, within $18^\circ$ of the objects collected in the flux-limited catalog of AGNs observed in X-rays (Swift-BAT-70~\cite{Baumgartner2013}) closer than 130~Mpc and brighter than $10^{44}~$erg/s. Note that the most significant excess (4.3$\sigma$ pre-trial) observed in the blind search corresponds to a region close to the super-Galactic plane and to the direction of Cen A, at a similar energy threshold, 54~EeV, and similar angular scale, $12^\circ$. A sky map in Galactic coordinates of the Li-Ma significances of overdensities in 12$^\circ$-radius windows for the events with $E\geq 54$~EeV is shown in figure~\ref{fig:cenA}.

\begin{figure}[!h]
  \centering
  \vspace{0.5cm} \includegraphics[width=7.cm]{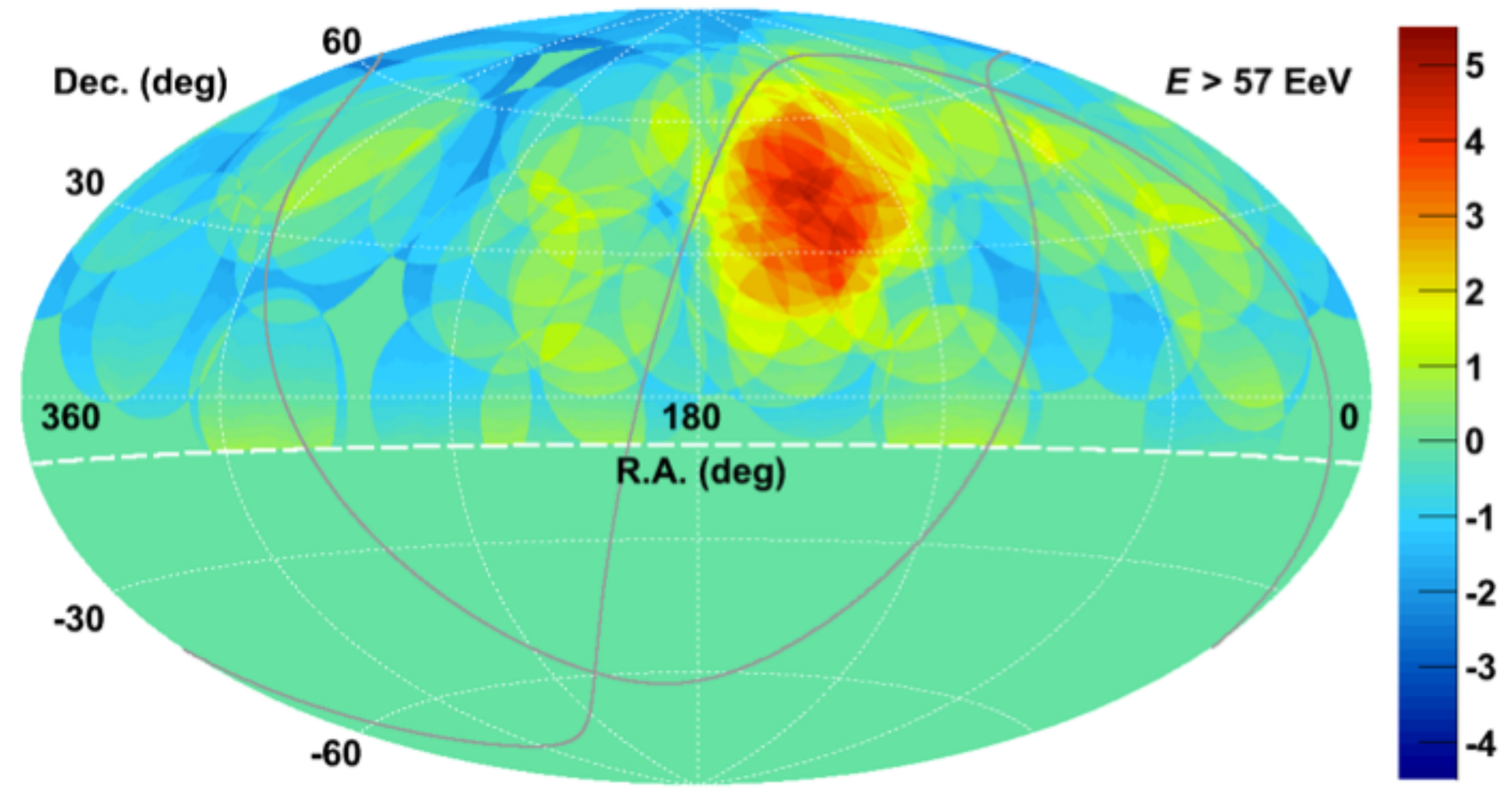}
  \caption{Sky map in equatorial coordinates of the Li-Ma significances of overdensities in 20$^\circ$-radius windows for the Telescope Array events with $E\geq 57$~EeV~\cite{TA2014}.}
  \label{fig:ta}
\end{figure}

At the Telescope Array, similar searches for anisotropies have also been performed above 57~EeV. Using data collected over a 5-year period, a cluster of events has been found by oversampling the sky map using 20$^\circ$-radius circles~\cite{TA2014}. The cluster of events, centered at about 19$^\circ$ off of the super-Galactic plane but on top of no known specific object, has a Li-Ma statistical significance of 5.1$\sigma$ (pre-trial). Once penalized for the angular scan performed each 5$^\circ$, the significance of the excess is 3.4$\sigma$. A confirmation of this excess requires more statistics. 

Overall, the absence of strong anisotropies at ultra-high energies is amenable to different interpretations. If the bulk of particles is made of light nuclei, that might indicate a large number of sources. If in turn the bulk of particles is made by large-$Z$ nuclei, then the lack of anisotropy might be caused by large deflections. Information on the mass of the primaries at these energies is thus also of relevance in the study of the distribution of the arrival directions.

\subsection{Multi-Messenger Approach}

In the context of testing the distribution of the arrival directions of the highest-energy CRs, it is worth mentioning the result of a full-sky study, conducted in a collaboration among Auger, Telescope Array and IceCube~\cite{MM2016}. It consists in the search for correlations between the directions of UHECRs observed at the Auger Observatory and at the Telescope Array, and those of very high-energy neutrino candidates detected by IceCube. Although no indications of correlations at discovery level are found, it is to be noted that the smallest post-trial $p-$values (corresponding to $\gtrsim 3\sigma$) are obtained when considering the correlations between the directions of cascade events observed by IceCube and those of the CRs. The excess of correlations, found at angular scales of $\simeq20^\circ$, arises mostly from pairs of events in the region of the sky where the Telescope Array has reported an excess of events and in regions close to the super-Galactic plane in correspondence with the largest excess observed in Auger data. Further insight will potentially arise from increased statistics and eventually with the inclusion of CR composition information that may become available so as to better model possible effects of magnetic field deflections. This may help to understand if there is a contribution in the astrophysical neutrino signal observed by IceCube correlated to the sources of the observed UHECRs.

\section{Conclusion}\label{sec:summary}

During the past decade, important observational results have been reported on the angular distributions of TeV-PeV CRs. While only dipolar excesses were expected, the myriad of reported anisotropies has led to important progresses on the understanding of the propagation regime of low-energy Galactic CRs. 

In contrast, the quest for finding UHECR sources is more difficult than expected a decade ago. Future work will profit from the increased statistics and ability to perform anisotropy searches with distinction based on the mass of the primaries as anticipated with the upgraded instrumentation at the Pierre Auger Observatory~\cite{Engel2015}. However, another jump in statistics appears necessary, keeping similar observable resolutions.

%\begin{acknowledgments}

%\end{acknowledgments}

\bigskip % extra skip inserted

\end{document}